\documentstyle[12pt,epsf]{article}
\newcommand{\ver}{H\gamma\gamma}
\newcommand{\im }{{\it Im}}

\begin{document}

\thispagestyle{empty}
\begin{flushright}
MZ-TH-95-18\\
BudkerINP-95-68\\
hep-ph/9508100\\
August 1995\\
\end{flushright}
\vspace{0.5cm}
\begin{center}
{\Large \bf Two -loop \boldmath{$O(G_F{M_H}^2)$} radiative corrections \\[.3cm]
to the  Higgs decay width \boldmath{$H \to \gamma \gamma $} for large \\[.3cm]
Higgs boson masses.}
\end{center}
\vspace{1.3cm}
\begin{center}
{J.G. K\"{o}rner\footnote{Supported by the BMFT, FRG, under contract
06MZ566}$^,$\footnote{Supported
in part by Human Capital and Mobility program,\\
\hbox{\qquad}EU, under contract CHRX-CT94-0579},
K. Melnikov$^{2,}$\footnote{Supported by Graduiertenkolleg
"Teilchenphysik", Mainz} and
O.I. Yakovlev$^{1,2,}$\footnote{Permanent address:
Budker Institute for Nuclear Physics, 630090, Novosibirsk, Russia}}\\[1cm]
{\bf \em Institut f\"{u}r Physik, THEP,  Johannes
Gutenberg Universit\"{a}t,}\\
{\bf \em Staudinger Weg 7, Germany, D 55099.}\\
\end{center}
\vspace{2cm}
\begin{abstract}
     This note is devoted to the calculation of the two-loop
$O(G_F {M_H}^2)$ radiative
corrections to the  Higgs decay width
$H \to \gamma \gamma $ for large values of the Higgs
boson mass $M_H$ within the Minimal Standard Model.
 The use of the Equivalence Theorem makes it possible to reduce
the problem to the
consideration of the physical Higgs boson field and the Goldstone bosons
$w^{+},w^{-},z$. We present analytical
results for the various two- and three-particle
absorptive parts of two-loop contributions,
using dispersive techniques, analytic results for all
but one of the dispersive contributions.
The typical size of the correction is $\sim ~30 $ percent  for a
Higgs boson mass of order $1~TeV$.
\end{abstract}

\newpage
{\large \bf 1.Introduction}
\par
The neutral scalar Higgs boson is the essential ingredient of the Standard
   Model of the electroweak interactions.
The Higgs boson mass is a free parameter in the Minimal Standard Model
and until now we do not know much about its value.
Experiments exclude a Higgs boson lighter then  $\sim~65$ GeV.
Also theoretical arguments based on perturbative unitarity
suggest that the upper bound on the Higgs boson mass is
$O(1)$ TeV \footnote{This statement is also supported by lattice
investigations \cite{Lat}}.

It is widely believed that the properties of the Higgs boson can be
investigated at the Next Linear Collider which  will be
able to operate in different modes ($e^+e^- , e^{+;-} \gamma , \gamma \gamma
 $). In particular $\gamma \gamma $ collisions are well
suited not only for the observation of the Higgs boson signal but also
for studying its properties (for a review see Ref. \cite{Ginz}).

  As is known for a long time the $H\gamma\gamma$ vertex
  serves as a ``counter''
of the particles with  masses larger than the Higgs boson mass:
   if these particles acquire masses due to the standard Higgs mechanism,
then they do not decouple from the Higgs boson and provide a constant
contribution to the $H\gamma \gamma$ vertex. Therefore, the
 $H \gamma \gamma$ vertex
 can provide  us, in
principle,   with unique information about the structure of the
  theory at energy scales unachievable for modern accelerators.

  A similar point also shows up in some other aspect: it turns out that
 the $H\gamma\gamma$ vertex is very sensitive to different anomalous
 couplings in the massive gauge boson sector of the Standard Model (SM).
 All these properties make the $H\gamma\gamma$ interaction vertex an
 extremely interesting object from the theoretical point of view.
 In order to exploit the
possibility to look for deviations from the SM predictions for the
$H\gamma\gamma$ vertex, one needs quite accurate predictions for
 this
    vertex within  the framework of the Minimal Standard Model.

  At the tree level the $\ver$ vertex is absent in the Standard Model.
  At the one-loop level the W-boson and the top quark contribute
  to the effective $H\gamma\gamma$
form factor.  This one-loop result can be found in the text books
  \cite {Okun}. Note for the time being
that the contributions of the $W$ and
$t$-quark loop to the $H\gamma\gamma$ vertex  have different
signs and hence tend
  to compensate each other. For realistic masses of the $W$-boson and
the top quark this compensation occurs for Higgs boson
  masses $\sim 600$ GeV.

  The QCD radiative corrections to the $\ver$ vertex were
calculated recently by several groups \cite {QCD}.
 These corrections are negligible below $t\bar t$ threshold and are large
above the threshold. As for the size
of the other SM radiative corrections, we do not know much
about them at present. Recently the
corrections of order $O(G_F {m_t}^2)$ were evaluated in the limit of a
small Higgs mass \cite{Dj}.
In this paper we consider the leading $O(G_F {M_H}^2)$
    SM radiative corrections
   in the limit of large Higgs boson masses. We show that this correction has
the same order of magnitude  but the opposite sign as the QCD correction
  in the
interval $ 0.5~TeV  < m_H <  1.5~TeV $ and blows up for larger Higgs boson
  masses.

The technical tool which results in great simplifications of the
calculations is the use of the Goldstone Boson
Equivalence Theorem (ET) \cite {ET}.

 The organization of the paper is as follows: in section $1$ we
discuss the one-loop calculation of the $H\gamma\gamma$ vertex
    in the framework of the ET; section 2 is devoted to the two-loop
calculation: we briefly discuss the renormalization procedure and present
   results for the imaginary and real parts of the $H\gamma\gamma$
vertex; in section 3
 we discuss our final results and make some concluding remarks.

\vspace{1cm}
{\large \bf 2. Lowest order $H\gamma\gamma$ vertex.}
\vspace{1cm}

The interaction of the Higgs boson with two photons can be described
   with the help of the effective Lagrangian:
\begin{equation}
L=\frac {\alpha }{4\pi v}F(s)F_{\mu\nu}F^{\mu\nu}H
\end{equation}

In this equation $F(s)$ denotes a form factor which contains all information
about the particles propagating in the loop. In the Minimal Standard Model
the form factor $F(s)$ obtains contributions from the top quark and
     $W$-boson .

\begin{figure}[htb]
\epsfxsize=8cm
\centerline{\epsffile{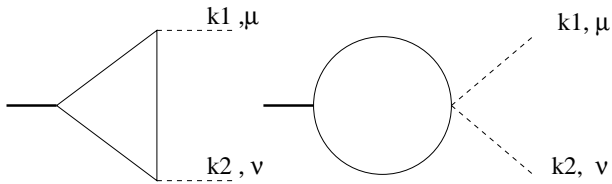}}
\caption[]{Generic lowest order graphs. The dashed lines correspond to photons,
heavy solid lines are Higgs bosons.
The particles inside the loop (light solid lines) are $W$ boson
and top quark.}
\end{figure}

The lowest order contribution to the $H\gamma\gamma$ vertex
is given by the graphs shown in Fig.1.
The analytical results for the fermion and spin-one boson contributions can
be found e.g. in \cite{Okun}.
In the limit when the Higgs mass is large in comparison with the
mass of the particle propagating in the loop, the contribution of the
fermions to $F(s)$ is
suppressed as $(\frac {M_f}{M_H})^2$ while the contribution of the $W$-loop
results in a constant :
\begin{equation}
{F^{(0)}}_{m_H \to \infty} \to  2   \label{flim}
\end{equation}

This asymptotic value can be obtained using the Goldstone Boson Equivalence
Theorem which states that in the limit of a large Higgs mass
$M_H \gg M_W $ the leading $O(G_F {M_H}^2)$ contribution to a given Green's
 function can be obtained by replacing the gauge bosons $W,Z$
by the corresponding would-be Goldsone bosons $w,z$ of the symmetry breaking
sector of the theory.
 The Goldstone bosons can be taken to be massless with the desired
 accuracy \cite {ET}.

 The interaction of the would-be Goldstone bosons with the Higgs and
photon fields is described by the $U_{EM}(1)$ gauged linear $\sigma $-model:
\begin{eqnarray}
L&=&{(D_{\mu}w)}^{*}{(D^{\mu}w)}+\frac {1}{2}\partial _{\mu}z
 \partial ^{\mu}z
+\frac {1}{2}\partial _{\mu}H \partial ^{\mu}H
  -\frac{1}{2}{M_H}^2 H^2\nonumber \\
&-&\frac {{M_H}^2}{4v^2}{({\Phi}^2+H^2)}^2-\frac {{M_H}^2}{v}(\Phi^2+H^2)H
-\frac {1}{4} F_{\mu \nu} F^{\mu \nu}
\end{eqnarray}

Here $D_{\mu} =\partial _\mu -ieA_{\mu} $
is the $U_{EM}(1)$ covariant derivative,
$M_H$ is the mass of the Higgs field, $v$ is its vacuum expectation value
and $\Phi $ is the triplet of the Goldstone bosons $w^+,w^-,z$.
The Feynman rules for this Lagrangian can be found e.g. in Ref. \cite{FR}.

  Let us first reproduce the result of Eq.(\ref{flim})
  using the Lagrangian of Eq.(3).
It is straightforward to write down the sum of the Feynman graphs shown
in the Fig.1 ( neglecting for the moment the contribution from the top loop).
 The contribution  to the form factor $F(s)$ can be conveniently
  obtained by contracting the one-loop tensor
amplitude with the tensor (the notations for outgoing photons are
clarified in Fig.1):
\begin{equation}
d^{\mu\nu}=g^{\mu\nu}k_1k_2-{k_1}^{\nu}{k_2}^{\mu}
\end{equation}
In spite of the fact that the sum of these graphs should be
ultraviolet finite, we need to regularize at intermediate steps of
the calculation. For simplicity we adopt dimensional regularization,
 working in $d$ dimensions.
At the end of the calculations we shall put $d$ equal to 4.
After some algebra one
finds for the sum of the lowest order amplitudes:
\begin{equation}
M=M_{\mu\nu}d_{\mu\nu}= {M_H}^2 2\pi\alpha
(d-4)s\int \frac {d^dq}{{(2\pi)}^d} \frac {1}{{(k_1+q)}^2{(k_2-q)}^2}
\end{equation}
{}From this equation it is seen that the  leading
order  calculation amounts to the calculation of
the divergent part of the massless two-point function.
Using well-known results for the two-point function in Eq.(5),
 we obtain the asymptotic result given in Eq.(\ref{flim}).

It is also possible to calculate these graphs using dispersion relations.
In order to do this, we need to cut the graphs shown on the Fig.1 in all
possible ways, calculate the contribution of the cut graphs to the
imaginary part of the $F(s)$ using unitarity relation  and
 finally integrate the imaginary part of the $F(s)$
along the cut.
As our Goldstone bosons are exactly massless, the cut goes from $0$ to
$\infty$ in the complex $s$-plane. If we cut the graphs of
Fig.1, the imaginary part of $F(s)$ is given by the
convolution of the decay amplitude $H(s) \to w^+w^-$ with the amplitude
 $ w^+w^- \to \gamma \gamma $. Note that conservation of the total angular
 momentum
requires equal helicities of both photons in the final state.

It is not difficult to see by exact calculation
    that the amplitude $w^+w^- \to \gamma
 \gamma $ vanishes for massless $w^+$ and $w^-$ bosons
in the equal photon helicity configuration.
Therefore the imaginary part of the
$F(s)$ is zero and one fails to reproduce the result of the
direct evaluation of the Feynman graphs.
To find a way out of this paradox we need to investigate the
amplitude $w^+w^- \to \gamma \gamma $ more carefully. For this aim we
introduce a mass for the Goldstone bosons which now serves as
an infra-red cut-off. The amplitude is then:
\begin{equation}
d^{\mu\nu}M_{\mu\nu}(w^+w^- \to \gamma \gamma )
=ie^2 \frac {2m^2s^2}{(t-m^2)(u-m^2)}
\end{equation}
where $m$ is the mass of the Goldstone bosons and $t$ and
 $u$ are the Mandelstam
variables of the process.

 It is then straightforward to calculate the
imaginary part of the $F(s)$ to the lowest order. One obtains
\begin{equation}
 {\it Im} F^{(0)}(s)=-\pi {{M_H}^2} \frac {4m^2}{s^2} \label{loim}
\log \left( \frac {1+\beta}{1-\beta}\right )
\end{equation}
where $\beta $ is the velocity of the (massive) Goldstone boson.
If we put the mass of the Goldstone boson equal to zero in Eq.(\ref{loim}),
  the
imaginary part of $F(s)$ is zero in accordance with the
previous statement. However, the lower limit in the
dispersion integral is $4m^2$.  In fact, if we consider
the imaginary part given by Eq.(\ref{loim}) in the dispersion integral, we
can see that in the limit $m \to 0$ the imaginary part of $F(s)$
turns into a $\delta$(s)-function.

  Hence,  the correct procedure consists in evaluating the
 dispersion integral with
finite Goldstone boson  masses
and taking the limit $m \to 0$ only after the
  integration over the cut has been performed.

In this way, we obtain the same result as in Eq.(\ref{flim})
for the real part of
$F(s)$, as has been obtained from the known complete expression
 for $F(s)$ in
the large Higgs mass limit or from the direct evaluation of the Feynman
graphs with massless Goldstone bosons.

  The reason why we have discussed the one-loop calculation
of the $H\gamma\gamma$ vertex in some detail is
two-fold: first, it serves as a reference point - to justify the use of the
 Equivalence Theorem for the two-loop calculation; second, in our opinion this
calculation shows some unexpected properties ( for instance, the
evaluation of this one-loop result through the dispersion relations is very
similar to the evaluation of the axial anomaly through the imaginary part
of the triangle graph \cite{Dolg}. However, we have failed to find deep reasons
   underlying  this similarity).

\newpage
{\large \bf 3.Two -loop contribution to the $H\gamma\gamma$ vertex}
\vspace{0.5cm}

{\bf 3.1 Renormalization}

   In this subsection we briefly discuss the renormalization
procedure which is needed for the evaluation of the two-loop graphs.
First note, that as the $H\gamma\gamma$ interaction is absent
 in the SM lagrangian, the two-loop graphs must be finite after we
renormalize all subdivergencies. In other words, to make our two-loop
 amplitude finite, we need only one-loop counter terms. The latter are
constructed according to the following procedure.

The "matter" part of the Lagrangian (Eq.3) contains two
independent parameters:
the mass of the Higgs field $M_H$ and the vacuum expectation value $v$.
We fix the one-loop counter-terms
by requiring the mass of the Higgs field and the
vacuum expectation value to be
exact one-loop quantities. This requirement eliminates all tadpole graphs
and provides us with the counter-terms for all other divergent subgraphs.
For instance, self-energies of the Goldstone bosons must
be effectively subtracted on mass-shell.
  Further we will need the counter-terms for the vertecies
$Hw^+w^-$ and $Hzz$ which can also be obtained from above requirements.

The next point is the renormalization of the $\gamma w^+w^-$
vertex. As  this vertex is convergent, its
renormalization is fixed by the renormalization of the
 Goldstone boson wave function
which in turn
is fixed by the renormalization of the self-energy operator for the
Goldstone boson. This procedure is compatible with the electromagnetic
Ward identities  of the gauged $\sigma$-model.

\vspace{0.5cm}
{\bf 3.2 Two-particle cuts.}

In this subsection we compute the contributions of the two-particle cuts
 of the graphs  presented in Figs. 2-5. The simplest (quasi one-loop)
 contributions are given by the set of Feynman
graphs shown in the Fig.2 and the two-particle cuts of the graphs in Fig.5.

\begin{figure}[htb]
\epsfxsize=10cm
\centerline{\epsffile{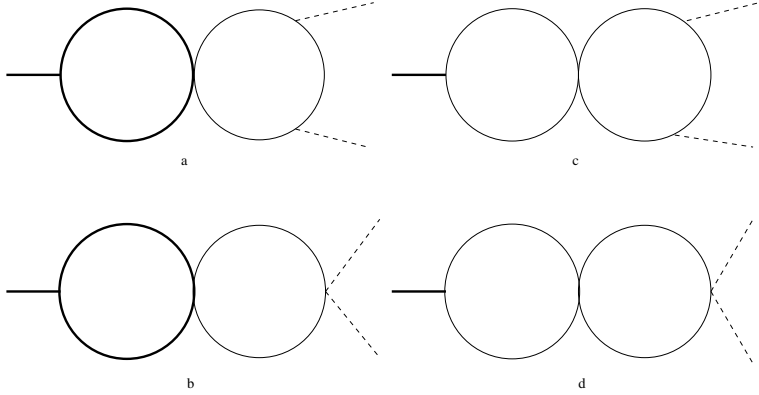}}
\caption[]{``Quasi one-loop'' two-loop diagrams. Heavy solid line denote Higgs
 bosons, thin solid lines denote $w^+$,$w^-$,$z$ goldstone bosons of the ET.
Dashed lines are photons.}
\end{figure}

 The graphs shown in the Fig.2 are quasi one-loop graphs. As
the $Hw^+w^-$ vertex diverges at the one-loop level one needs
  to bring in counter-terms
which can be obtained according to the recipe given above. It is
also convenient to consider simultaneously the graphs shown in Fig.2 and
the two-particle cuts of the two-Higgs-two Goldstone boson
vertex graphs presented in Fig.5. Summing up
 the
contributions
of Fig.2, the two-particle cut contributions of the
 graphs shown in Fig.5 and the one-loop
 counter-term for the $Hw^+w^-$ vertex, one has:

\begin{equation}
{F^{(1)}}= 2{\left (\frac {M_H}{4\pi v} \right )}^2
(2-\sqrt {3} \pi)
\end{equation}

  Next we discuss the contribution of the graphs shown in
Fig.3. Note that we are considering only two-particle cuts in this
section.

\begin{figure}[htb]
\epsfxsize=10cm
\centerline{\epsffile{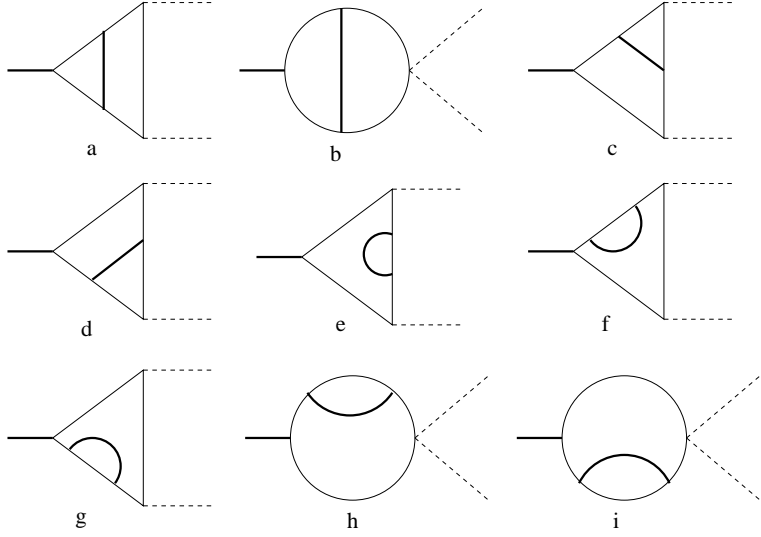}}
\caption[]{Abelian (QED-like) two-loop diagrams,
 Line drawings as in Fig.2.}
\end{figure}

Let us start with the graphs shown in Fig.3(a-b) and
consider the two-particle cuts which lie to the right of the virtual
Higgs boson line. The cut  contribution is given by the convolution of
 the one-loop
$H \to w^+w^-$ amplitude (with the Higgs boson in the t-channel)
with the Born amplitude for
$w^+w^- \to \gamma \gamma$. As we know from the discussion of the lowest
order vertex, the latter is singular for small values of $s$.
Unfortunately, the one-loop correction to $H \to w^+w^-$ is also singular
for $s=0$ if the Goldstone bosons are exactly massless. As before  we have to
introduce a mass $m$ for the Goldstone boson to handle this infrared
divergence. Note at this point, that the ET theorem,
 guarantees the existence of a smooth limit as $m \to 0$. Hence, we expect
that the sum of all two-loop contributions will not be sensitive to
 the details of the infrared limit of the theory.

Evaluating the $Hw^+w^-$ vertex in the limit $m_H >> \sqrt {s}, m$ we
find the following result:
\begin{equation}
F _L = \frac {{M_H}^2}{2} {\left (\frac {M_H}{4\pi v}\right )}^2
\left (1+\log \left(\frac {{M_H}^2}{m^2}\right)
 -\beta \log \left (\frac {1+\beta}{1-\beta} \right ) \right )
\end{equation}

Putting everything together the imaginary part corresponding to
 the "right cut"
graphs of Fig.3(a,b) is given by:
\begin{equation}
  {\it Im}F^{(2)}=-\frac {{M_H}^2}{2} \cdot 2\pi \frac {4m^2}{s^2}
\log \left (\frac {1+\beta}{1-\beta}\right ) F _L(s)
\end{equation}

Inserting Eq.(10) into a dispersion integral we can evaluate the
contribution of these cut graphs to the real part of the $F(s)$ and
get:
\begin{equation}
F^{(2)} = 2 {\left (\frac {M_H}{4\pi v}\right )}^2
\left (1+\log \left (\frac {{M_H}^2}{m^2}\right )
 -\frac {4}{3}\left (1+\frac {\pi ^2}{12}\right )\right )
\end{equation}

  Another possibility to cut the graphs Fig.3(a,b) is to cut
to the
left of the virtual Higgs line.
We divide the integration region in the
dispersion integral into two parts introducing an arbitrary scale $\mu $.
   The scale $\mu$ can be chosen to satisfy the following inequalities:
      $$ m \ll \mu \ll M_H.$$
If we are interested in the contribution from the ``high --
energy'' part of this graph, we can put the masses of the Goldstone bosons
equal to zero. For the ``high--energy'' part of the imaginary part of
$F(s)$ we obtain:
\begin{eqnarray}
\im {F^{(3)}}&=&-\frac {{M_H}^4}{2}{\left (\frac {M_H}{4\pi v}
\right )}^2 \frac {\pi}{s^3}~A \nonumber \\
 A =
 \Bigg ( 8s-4 \Big (s  &+&
  {M_H}^2 \Big )\log \left (\frac {s+{M_H}^2}{{M_H}^2}\right )
 + 4{M_H}^2
Li_2 \left (-\frac  {s}{{M_H}^2}\right ) \Bigg )
\end{eqnarray}
  Inserting this expression into the
dispersion integral we can evaluate the contribution
  of
the "high-energy" part to the real part of $F(s)$, where we must
remember
that the lower limit for the integration of the above quantity
is given by $\mu$.

   Performing the integration, we  get:
\begin{equation}
F^{(3)}= 2 {\left (\frac {M_H}{4\pi v}\right )}^2
\left (\frac {1}{4} \log \left (\frac {\mu ^2}{{M_H}^2}\right )- \frac {7}{2} +
 \frac {\pi ^2}{6} +\frac {3}{2} \zeta (3)\right )
\end{equation}

Next we have to find the contribution of the
  ``low-energy'' region of this graph to the $F(s)$.  We do this
  by expanding
     the amplitude in terms of powers of
 $\frac {\sqrt {s}}{M_H}$ and $\frac {m}{M_H}$.

The result for the imaginary part reads:
\begin{equation}
{\it Im}F^{(3)}
 =-\pi {{M_H}^2}{\left (\frac {M_H}{4\pi v} \right )}^2 \frac {\beta}{s^2}
\left (- \frac {s}{2} +4m^2\left
(\frac {\pi^2}{2}-\log ^2 \frac {1+\beta}{1-\beta}
 \right ) \right )
\end{equation}

Inserting (14) into the dispersion integral and integrating
from $4m^2$ up to $\mu^2$ we find the ``low-energy'' contribution
to the real part of $F(s)$:
\begin{equation}
F^{(3)}=2{\left (\frac {M_H}{4\pi v}\right )}^2
\left (-\frac {1}{4} \log \frac {\mu ^2}{m^2}-\frac {5}{6}
 +\frac {\pi ^2}{18}\right )
\end{equation}

  Finally  we have to sum
the "low-energy" and "high-energy" contributions and get:
\begin{equation}
F^{(3)} = 2{\left (\frac {M_H}{4\pi v}\right )}^2
\left (-\frac {1}{4} \log \frac {{M_H}^2}{m^2}
-\frac {13}{3} +\frac {\pi^2}{9}+\frac {3}{2} \zeta (3)\right )
\end{equation}

  The next two-particle cut contributions that we have to consider are
 obtained by cutting the
  graphs presented in Fig.3(c,d,e).
   The calculation  proceeds in
complete analogy with the case considered in details above.
  The result of our evaluation is:
\begin{equation}
F^{(4)}= 2{\left (\frac {M_H}{4\pi v}\right )}^2
\left (-\frac {3}{4} \log\frac {{M_H}^2}{m^2} -\frac {\pi^2}{2}+
\frac {3}{2} \zeta (3) +3 \right )
\end{equation}
We mention that the graphs Fig.3(f,h,g,i) have no two-particle cuts due
to on-shell renormalization
of the Goldstone bosons.

If we sum $F^{(2)}$, $F^{(3)}$ and $F^{(4)}$ we see
 that the sum is finite in the limit $m \to 0$ in
agreement with our expectations:
\begin{equation}
F^{(2)}+F^{(3)}+F^{(4)}
 = 2 {\left (\frac {M_H}{4\pi v} \right )}^2
\left (3\zeta (3) -\frac {\pi ^2}{2} -\frac {5}{3} \right ) \label{TPFig4}
\end{equation}
To recapitulate, Eq.(18) contains the contribution of the two-particle
 cuts of the diagrams in Fig.3.

Next we are going to discuss the two-particle cut
 contribution corresponding to graphs presented
in Fig.4(a-b). Similar to the
situation discussed above there are two possible ways of cutting
these graphs, i.e. to the
left and to the right of the virtual Goldstone boson line.

\begin{figure}[htb]
\epsfxsize=10cm
\centerline{\epsffile{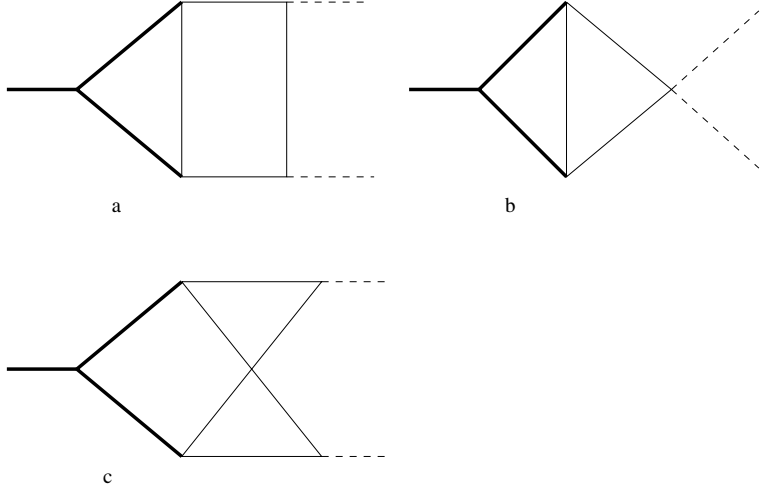}}
\caption[]{Two loop diagrams with triple Higgs coupling.
Line drawings explained in Fig.2}
\end{figure}

We start with the contribution of the right-cut graph.
Its contribution is given by the
convolution of the correction to the $Hw^+w^-$ vertex and the
    $w^+w^- \to \gamma \gamma$
amplitude. In this case the $Hw^+w^-$ vertex is not singular for $s=0$
when the Goldstone bosons are massless. Hence the  contribution of this
``right-cut'' graph is simply given by the product of the lowest
order $w^+w^-\to \gamma \gamma$ and the $Hw^+w^-$ vertex calculated for $s=0$.
One obtains:
\begin{equation}
F^{(5)}=2 {\left (\frac {M_H}{4\pi v} \right )}^2 \cdot 3
\end{equation}

The contribution of the ``left-cut'' graphs  is also
calculated straightforwardly \footnote { There is one subtlety in this
discussion.
Considering this cut graph more carefully we find both real
 and {\bf imaginary} parts originating e.g. from the
imaginary part of the box graph $HH \to \gamma \gamma$. For our purposes we
need only the real part of the amplitude, which ( after being integrated over
the intermediate particle phase space in the unitarity relation) results in
Eq.(20). As for the {\bf imaginary} part of the box graph, it will be
exactly canceled by the imaginary part of the {\bf three particle cut}.
 In the latter case, the imaginary part comes from the pole of the virtual
Higgs boson propagator, which
comes into play when
total energy of the
process is larger than $2M_H$ ( see also the discussion after  Eqs.(38) and
(39)).}.
After a little algebra we find the following
result for the imaginary part :
\begin{equation}
 \im F^{(6)}= \frac {3\pi}{2} {\left (\frac {M_H}{4\pi v} \right )}^2
 \frac {{M_H}^4}{s^2}
 2\beta _H \left ( \frac {1+{\beta _H}^2}{\beta _H}
\log \left (\frac {1+\beta _H}{1-\beta _H} \right )-2 \right )
\end{equation}
In this equation
    $$\beta _H = \sqrt{1-\frac {4{M_H}^2}{s}}$$
is the velocity of the Higgs boson
in the intermediate state. Note, that the dispersion integral
       starts at the point $s=4{M_H}^2$. The result of the integration
is given by:
\begin{equation}
F^{(6)}=2 {\left (\frac {M_H}{4\pi v} \right )}^2 \left (\sqrt {3} \pi
  -\frac {\pi^2}{6}-\frac {15}{4} \right )
\end{equation}

The next step is the evaluation of the contribution of the graph presented
in
Fig.4(c). There is only one possibility to obtain a
two-particle cut from this graph -- it is the cut with the two Higgs bosons
in the intermediate state. The evaluation of this cut is much more
involved due to its non--planar topology. Some details
 of our evaluation of this graph are given below.

First, after cutting the graph, we face the necessity to evaluate the
box graph. Contracting the box amplitude with the $d_{\mu\nu}$ tensor
 (defined in Eq.(4)), we
find the following representation for the box graph contribution:
\begin{equation}
T_{\mu\nu}d_{\mu\nu}= \frac {i}{{(4\pi)}^2} \int dydz \left ( \frac
{2s}{g(y,z)}
+ \frac {-sm^2+({M_H}^2-t)({M_H}^2-u)}{g^2(y,z)} \right )
\end{equation}
where $s,t,u$ are the usual Mandelstam variables and the function $g(y,z)$
reads:
$$ g(y,z)= m^2+(u-m^2)y+(t-m^2)z+syz $$
The $y$ and $z$ integrations in Eq.(22) extend from $0$ to $1$.
After integration over $y$ and $z$ we get the following
result for the box graph amplitude:
\begin{eqnarray}
M=4\pi \alpha \cdot i \left ( \frac {M_H}{4\pi v} \right )^2 2 \left(
\log ^2 \left (\frac {tu-{M_H}^4}{(t-{M_H}^2)(u-{M_H}^2)} \right )
\right. &-& \left. 4Li_2(1)+ \right. \nonumber \\ \left.
 2Li_2\left (\frac {tu-{M_H}^4}{(t-{M_H}^2)(u-{M_H}^2)} \right )
 \right. &+& \left.
  Li_2 \left (-\frac {{(t-{M_H}^2)}^2}{tu-{M_H}^4} \right )+
  \right. \nonumber \\ \left.
 Li_2 \left (-\frac {{(u-{M_H}^2)}^2}{tu-{M_H}^4} \right )
 \right. &+& \left.
 \frac {1}{2} \log \left (\frac {-tu}{{M_H}^4} \right ) \right)
\end{eqnarray}

In this equation $Li_2(x)$ is a Spence function as defined in reference
 \cite {Lew}.
To calculate the contribution of the box to the imaginary part of
the formfactor we have to integrate Eq.(23) over two particle
   phase space. In doing so, it is convenient to introduce a new
 variable $0< x <1$ according to:
\begin{equation}
\frac {s}{{M_H}^2} = \frac {(1+x)^2}{x} \label {x}
 \end{equation}
Then, the contribution of the box to the imaginary part of the formfactor
is given by:
\begin{eqnarray}
\im F^{(7)}=-2 \cdot \frac {3}{2} \left (\frac {M_H}{4\pi v}
\right ) ^2
\frac {{M_H}^4}{s^2} \Bigg ( \frac {1-x}{1+x}  \Big (
 2\log ^2(x) &-&
2\frac {1+x^2}{1-x^2} \log (x) -2{\pi}^2-2  \Big ) + \nonumber \\
   8 Li_2(-x) +
2\frac {{\pi}^2}{3} -2 \log ^2(x) &+&  8 \log(x) \log(1+x)
 \Bigg )
\end{eqnarray}

Finally, in order to obtain its
contribution to the real part of $F(s)$ one needs to integrate the
imaginary part along the cut. It is clear from the graphs Fig.4(c) that
the cut goes from $4{M_H}^2$ to $\infty$. Performing this calculation we
find:
\begin{equation}
F^{(7)}=-2 {\left (\frac {M_H}{4\pi v} \right )}^2
\frac {3}{2} \left (-\frac {4}{9} \zeta (3) +\frac {38}{27\sqrt {3}} \pi ^3-
\frac {23}{9} \pi ^2 + \frac {2}{\sqrt {3}} \pi -\frac {11}{6} +8 C_1
\right )
\end{equation}
 Here  the constant $C_1$ is :
\begin{equation}
C_1= \int _{0}^{1} dx \frac {\log (x)\log (x^2+x+1)}{1+x} = -0.194692
\end{equation}
The result Eq.(26) completes the list of the two-particle cut
contributions.

\vspace{0.5cm}
{\bf 3.3 Three-particle cuts.}

    This subsection is devoted to the discussion of the three-particle
  cuts. First, we consider the graphs
corresponding to Fig.3(f,g,h,i). We remind the
reader that
 these graphs have no two-particle cuts due to the on-shell renormalization
 of the Goldstone bosons. In order to evaluate the three-particle
 intermediate state contribution, we have
to consider the convolution of the two
 processes $H \to w^+w^-H$ and $(Hw^+)w^- \to \gamma \gamma $.
As indicated by bracket the latter
 process can be viewed as the annihilation of the massless particle $w^-$
 and the massive particle $(Hw^+)$ into two photons. It is not difficult
  to
 calculate the $d^{\mu \nu}$-contracted
amplitude for $(Hw^+)w^- \to \gamma \gamma $
which reads:
\begin{equation}
M^{\mu\nu}\left ((Hw^+)w^- \to \gamma \gamma \right )
 d_{\mu\nu}= ie^2 \frac {{M_H}^2}{v} \label {simp}
 \end{equation}
It is then clear that the problem of the calculation of the imaginary part for
this cut contribution amounts to the problem of averaging the virtual Goldstone
boson
propagator on the left side of this graph over three-particle phase space.
 Performing the integration we find:
 \begin{equation}
\im F^{(8)}= -\frac {{M_H}^4}{2} {\left (\frac {M_H}{4\pi v}
  \right)}^2 \frac {2\pi}{s^3}
\left (-2(s-{M_H}^2)+(s+{M_H}^2)\log \left (\frac {s}{{M_H}^2}\right )
\right )
 \end{equation}

 We finally substitute this expression into the dispersion integral and
integrate along the cut going from $s={M_H}^2$ to $s= \infty $.
The result of this integration is:
 \begin{equation}
 F^{(8)}= 2{\left (\frac {M_H}{4\pi v} \right )}^2 \left (\frac {13}{8} -
 \frac {\pi ^2}{6} \right )
 \end{equation}

Next we discuss the three-particle cuts of the graphs Fig.3(a,b).
Cutting these graphs along the three particle intermediate
state contributions, it is easy
to see that these graphs produce exactly the same result as
the graphs discussed previously (Fig.3 f,h,g,i).

A more non-trivial situation arises for the three-particle cut of the graphs
shown on Fig.3(c,d). In this case the complexity stems from the fact
 that the amplitude to the right
of the cut does not have a simple form as in
Eq.(\ref {simp}). The way we proceed is the
following: as before we first contract this
amplitude with the tensor $d_{\mu\nu}$, and then perform the phase space
integration over the momentum of the decay products of the virtual
Goldstone boson. Then we obtain the following representation for the
imaginary part of $F(s)$:
\begin{equation}
\im F^{(9)}=-\frac {8\pi ^2M_H^2}{s^2} {\left (\frac {{M_H}^2}{v}
\right )}^2
\int \frac {d^3p_2}{{(2\pi )}^3 2E_2} \frac {\Gamma _2 (Q)}{Q^2}
\left (\frac {s}{Qk_1} -1 \right )
\end{equation}
where $ Q=k_1+k_2-p_2$ and $\Gamma _2 (Q)$ is given by
$$ \Gamma _2 (Q) =\frac {1}{8\pi } \frac {Q^2-{M_H}^2}{Q^2} $$
Integrating Eq.(31) we obtain the following result for the
 contribution of this graph to the imaginary part of $F(s)$:
\begin{eqnarray}
\im F^{(9)}&=& -2 \pi {\left (\frac {M_H}{4\pi v}\right )}^2
 \frac {{M_H}^4}{4s^3}~A \\
 A=
\left( 2s\log ^2 \left (\frac {s}{{M_H}^2} \right )-
 (6s \right. &+& \left. 2{M_H}^2)\log \left (\frac {s}{{M_H}^2} \right ) +
 8(s-{M_H}^2) \right ) \nonumber
\end{eqnarray}

\begin{figure}[htb]
\epsfxsize=10cm
\centerline{\epsffile{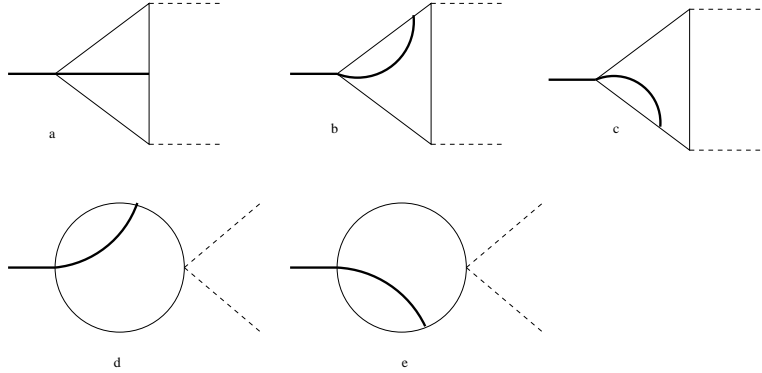}}
\caption[]{Two-loop diagrams with  two Higgs -- two Goldstone boson
interaction vertices.}
\end{figure}

  Integrating Eq.(32) along the cut we finally obtain
the contribution to the real
part of $F(s)$ which reads:
 \begin{equation}
 F^{(9)}=2 {\left (\frac {M_H}{4\pi v} \right )}^2 \frac {1}{4}
\left (\frac {4\pi ^2}{3} -\frac {17}{2} -4\zeta (3) \right )
\end{equation}

Next we come to the discussion of the graphs shown in Fig.5.
The calculation is performed in complete analogy with the case discussed
previously. Without going into details we present the result for
the imaginary and real parts of the corresponding cut graphs:
The joint contribution of the cut graphs Fig.5(b,d) to the imaginary part is
the same as the contribution of the cut graphs Fig.5(c,e) and it has the form:
\begin{equation}
\im F^{(10)}= -{2\pi} {\left (\frac {M_H}{4\pi v} \right )}^2
 \frac {{M_H}^2}{2s^3}
 \left (\frac {(s+{M_H}^2)(s-{M_H}^2)}{2}
 -s{M_H}^2\log \left (\frac {s}{{M_H}^2} \right )
 \right )
\end{equation}
Upon integration we get for the real part:
 \begin{equation}
 F^{(10)}=2 {\left (\frac {M_H}{4\pi v} \right )}^2
\left (\frac {\pi ^2}{12} -\frac {7}{8} \right )
\end{equation}

 Next, let us write down the contribution of
 the cut graph shown in Fig.5a:
\begin{eqnarray}
\im F^{(11)}&=&-2\pi {\left (\frac {M_H}{4\pi v} \right )}^2
 \frac {{M_H}^2}{4s^3}~B \\
B= \left( -2s{M_H}^2 \log ^2 \left (\frac {s}{{M_H}^2} \right )
 \right. &-& \left. 2s{M_H}^2\log \left (\frac {s}{{M_H}^2} \right )
+(s-{M_H}^2)(3s-{M_H}^2) \right ) \nonumber
\end{eqnarray}
 Correspondingly, one has for the real part:
 \begin{equation}
F^{(11)}=2 {\left (\frac {M_H}{4\pi v} \right )}^2 \frac {1}{4}
\left ( 4\zeta (3)+\frac {\pi ^2}{3} -\frac {17}{2} \right )
 \end{equation}

 Next we consider the contribution of the three-particle cuts of the graphs
shown in  Fig.4(a,b). The first step of the calculation is similar to the
evaluation of the graphs Fig.5(b-e) because the right-hand side of the cut
graph is again given by the simple expression Eq.(\ref {simp}). Performing
all further integrations over the phase space variables, we obtain the
following representation for the imaginary part of the sum of these
cut graphs:
\begin{equation}
 \im F^{(12)}= -{2\pi}
 { \left (\frac {{M_H}^2}{4\pi v} \right )}^2
\frac {6{M_H}^4}{s^2} \int _{m}^{E_{max}} dE
 \frac {\sqrt {E^2-{M_H}^2}}{s-2\sqrt {s}E}
\end{equation}
where $E_{max}$ is given by
\begin{equation}
E_{max} = \frac {s+{M_H}^2}{2\sqrt {s}}
\end{equation}

One integrates over the energy of the virtual Higgs boson which
decays to two Goldstone bosons. The specific feature of this integral is
that, depending on the total energy of the process $ \sqrt {s}$, the
denominator of the integral can go through zero reflecting the fact that
 an intermediate state with two "real" Higgses can be
formed for $ s > 4{M_H}^2$. It is also clear
that for our purposes we have to treat
this singularity in the principal value sense.

It is straightforward to calculate this integral and one obtains
 the following
 expression for the imaginary part:
\begin{equation}
\im F^{(12)}=-2\pi {\left (\frac {{M_H}^2}{4\pi v} \right )}^2
\frac {6{M_H}^4}{s^2} \frac {1}{4} \left (-\frac {1}{2} \log\frac {s}{{M_H}^2}
-1+ \frac {{M_H}^2}{s}+\Psi (s) \right )
\end{equation}
where the function $\Psi (s) $ is defined by:
 \begin{equation}
\Psi (s)= \theta(4{M_H}^2-s)\frac {3}{2} ctg \left (\frac {\varphi}{2} \right )
  (\varphi-\frac {\pi}{3})
  -\theta(s-4{M_H}^2)\frac {3}{2}\frac {1-x}{1+x} \log (x)
\end{equation}
The variable $x$ is defined as in Eq.(\ref {x}) and $\varphi$
is defined through the relation
$$ s=4{M_H}^2 sin^2\frac {\varphi}{2} $$

Integrating the imaginary part we finally obtain the
following contribution to the real part of $F(s)$:
\begin{equation}
 F^{(12)}= 2 {\left (\frac {M_H}{4\pi v}\right )}^2
 \left (\frac {\pi ^2}{8} +\frac {3}{2} -
 \frac {3 \sqrt {3}}{2} Cl_2 \left (\frac {\pi }{3} \right ) \right)
\end{equation}
where $Cl_2(\varphi)$ is Clausen's function (see e.g. \cite {Lew}).

Now we are in the position to discuss the most difficult part of the
calculation, namely the evaluation of the contribution of
the three-particle cut given by the non-planar graph of Fig.4(c).
Performing the integration over the phase space, we obtain the
following representation for the contribution of this cut to the imaginary
part of $F(s)$:
 \begin{equation}
\im F^{(13)}=-{2\pi} {\left (\frac {{M_H}^2}{4\pi v} \right)}^2
\frac {6{M_H}^4}{s^2} (W_1(s)+W_2(s)+W_3(s))
\end{equation}
where $W_1, W_2$ and $W_3$ stand for:
\begin{eqnarray}
W_1(s) =2\left (2\log(2)+\frac {1}{2} \right )\int _{m}^{E_{max}}
dE \frac {\beta _H E}{s-2\sqrt {s} E} \\
W_2(s) =-2\int _{m}^{E_{max}} \frac {dE}{\sqrt {s}}
\frac {s-E\sqrt {s}}{s-2\sqrt {s} E}
\log \left (\frac {s-E\sqrt {s} +\beta _H E \sqrt {s}}
{s-E\sqrt {s} -\beta _H E \sqrt {s}} \right )\\
W_3(s) =-2\int _{m}^{E_{max}}
dE \frac {\beta _H E}{s-2\sqrt {s} E}
\log \left (\frac {(s-E\sqrt {s})^2}{(\beta _HE\sqrt {s})^2} -1 \right )
\end{eqnarray}

In this expression $E_{max}$ is defined through Eq.(39) and $\beta_H$ is
 the velocity of the Higgs boson:
$$ \beta _H = \sqrt {1-\frac {4M_H^2}{E^2}}$$
 In each of the above integrals there is a pole in the integrand
for total energies larger than two Higgs bosons masses. We first
evaluate each of these integrals in the case when $s > 4{M_H}^2$
and then perform an analytic continuation to the region
$s< 4{M_H}^2$.
 We shall not
present explicit expression for the imaginary part of this functions
 above threshold. If needed,
 it can be
obtained directly from the integral representation of the above
functions. One obtains:
\begin{eqnarray}
 W_1(s)+W_3(s)= \frac {1}{4}(1-\frac {{M_H}^2}{s})
 -\frac {1}{8}\log \left (\frac {s}{{M_H}^2} \right )
 &+&\frac {1}{8}\log ^2 \left (\frac {s}{{M_H}^2} \right )
 - \nonumber \\ \frac {{M_H}^2}{2s}\log \left (\frac {s}{{M_H}^2} \right ) +
\frac {1-x}{2(1+x)} \cdot
 \left ( -\frac {2\pi ^2}{3} -3Li_2(x)\right. &-& \left.
 2Li_2(-x) +\frac {7}{4} \log ^2(x)
 - \right. \nonumber \\ \left.
 3\log (x)\log (1-x^2)-\frac {3}{4}\log (x) \right. &+& \left.
 \frac {3}{2} \log (x)\log \left (\frac {s}{{M_H}^2} \right ) \right )
\end{eqnarray}
 \begin{eqnarray}
 W_2(s)= \frac {\pi ^2}{6}
 +\frac {{M_H}^2}{2s} \log \left ( \frac {s}{{M_H}^2} \right )
 -\frac {1}{2}\left (1-\frac {{M_H}^2}{s} \right )
 +\frac {1}{8} \log ^2 \left (\frac {s}{{M_H}^2} \right )
 -\frac {3}{8}\log ^2(x)
 \end{eqnarray}
Here again the variable $x$ is defined by Eq.(\ref {x}).

Performing the analytic continuation to the region $s<4{M_H}^2$
we find the following expressions for the above integrals:
\begin{eqnarray}
 W_1(s)+W_3(s)&=& \frac {1}{4} \left (1-\frac {{M_H}^2}{s} \right )
 -\frac {1}{8} \log \left (\frac {s}{{M_H}^2} \right )
 +\frac {1}{8}\log ^2 \left (\frac {s}{{M_H}^2} \right )- \nonumber \\
 \frac {{M_H}^2}{2s}\log \left (\frac {s}{{M_H}^2} \right ) &+&
\frac {ctg \left (\frac{\varphi}{2} \right )}{2}
 \Bigg ( (\varphi - \frac {\pi}{3}) \bigg ( \frac {3}{4}
-\frac {3}{2}\log \frac {s}{{M_H}^2}
 +3\log (2sin(\varphi )) \bigg ) + \nonumber \\
\frac {1}{2} \Big ( 3Cl_2(2\varphi )  &-&
   2Cl_2(\varphi ) \Big ) \Bigg )
 \end{eqnarray}
 \begin{eqnarray}
 W_2(s)= \frac {{M_H}^2}{2s}\log \left (\frac {s}{{M_H}^2} \right )
 -\frac {1}{2}\left (1-\frac {{M_H}^2}{s} \right )
+\frac {1}{8}\log ^2 \left (\frac {s}{{M_H}^2} \right )
 + \frac {3}{8} \left (\varphi -\frac {\pi}{3} \right )^2
 \end{eqnarray}
 Equations (47-50)  provide us with the desired result for the
contribution of this graph to the imaginary part of $F(s)$.

  The integration in the dispersion integral has to be done numerically and
we find:
\begin{equation}
 F^{(13)} = -2 {\left (\frac {M_H}{4\pi v} \right )}^2K,~~~~~K=0.0678
\end{equation}
  Summing up all contributions to the real part of $F(s)$  and
taking into
account permutations of the photon's legs where necessary, we obtain as
a final result:
\begin{eqnarray}
 F&=&F^{(0)}+\sum _{i=1}^{6}F^{(i)}+ 2\cdot F^{(7)}+4\cdot F^{(8)}+
    \nonumber \\
  &+&4\cdot F^{(9)}+2\cdot F^{(10)}+2\cdot F^{(11)}+
2\cdot F^{(12)}+4\cdot F^{(13)}\nonumber \\
  &=& 2 \cdot \left (1-3.027{\left (\frac {M_H}{4\pi v} \right )}^2
\right)
\end{eqnarray}
The result Eq.(52) completes our calculation and presents the two-loop
correction
to one-loop result Eq.(2). \\

\vspace{0.5cm}
{\large \bf 4. Discussion and conclusions}
\vspace{0.5cm}

Let us finally discuss the phenomenological implications of our
results. We begin with a discussion of the one-loop contribution
to $F(s)$ in Fig. 6.
\begin{figure}[htb]
\vspace{8cm}
\hspace{3.5cm}B\hspace{3cm}A\\
\vspace{-8cm}
\epsfxsize=10cm
\centerline{\epsffile{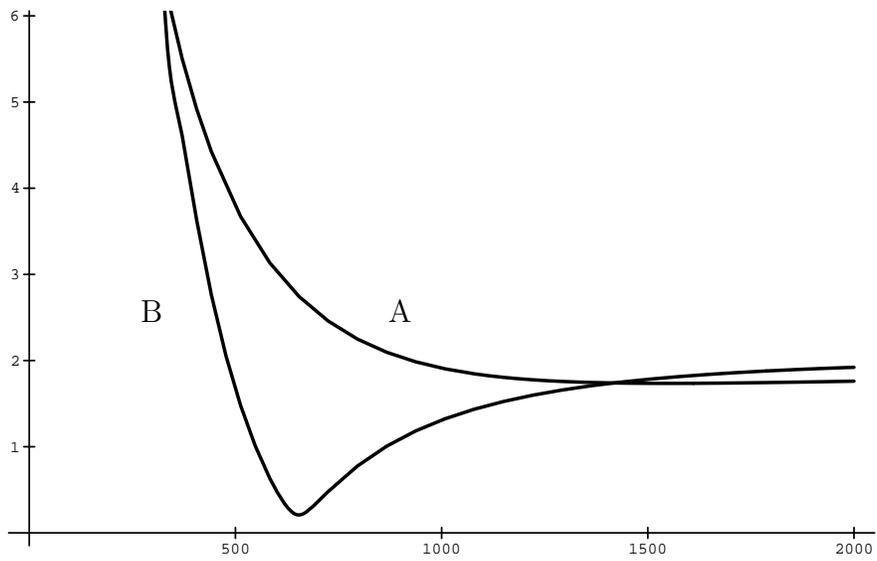}}
\caption[]{Absolute value of the one-loop form factor of process $H\to
\gamma \gamma$, $F(m_H)$, as a function of Higgs mass $M_H$ (GeV).
Curve A shows contribution of the W-boson only, whereas
curve B is the sum of the top
quark and the W-boson contributions.  }
\end{figure}
  As mass parameters we have taken  $m_t=180~GeV,~~m_W=80~GeV $.
Curve A shows a contribution of the W-boson only, whereas
curve B is the sum of the top
quark and the W-boson contributions.
Fig.6 shows that:
\begin{itemize}

 {\item
   the contribution of the W-boson to $F(s)$ is slowly approaching its
      asymptotic value  at $m_H>600~GeV$}

  {\item
  the contribution of the top quark is important
   until the Higgs mass reaches the value $m_H \sim 1~TeV$.
     We emphasize that there is a strong
 cancellation  between the contributions of the top quark and W boson
   for Higgs masses of order $m_H \sim 600~GeV$.}

\end{itemize}

As a consequence of these observations we expect that,
in the two-loop case, the use of the
 Equivalence Theorem for an estimation of the EW radiative corrections
   for the coupling of the Higgs boson to two photons
is reasonable  for the Higgs masses above $600~GeV$.
However, in order to make quantitative predictions in the region
$m_H \sim 600 GeV $
it is important to take into account the contribution which is proportional
top quark Yukawa coupling.

\begin{figure}[htb]
\epsfxsize=10cm
\centerline{\epsffile{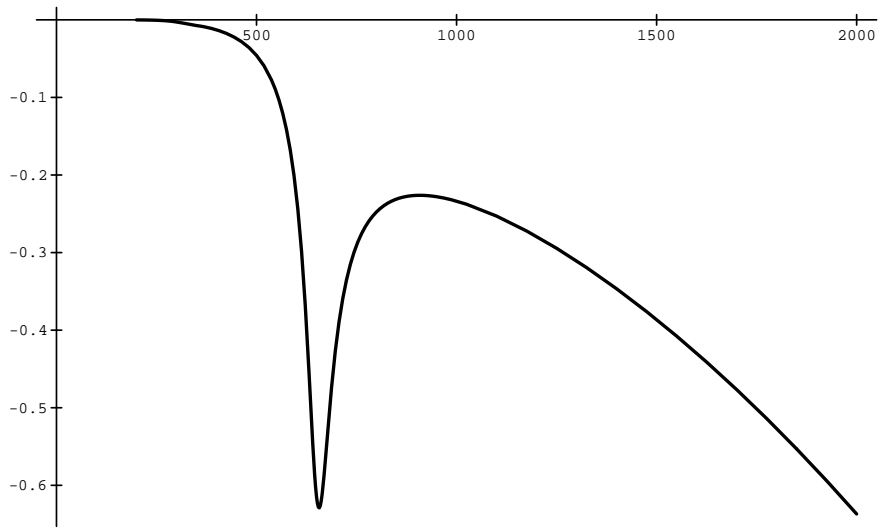}}
\caption[]{Relative two-loop electroweak correction to the decay width $H\to
\gamma \gamma$ (in percent) as a function of $M_H$ (GeV).}
\end{figure}

Our numerical results for the leading two-loop EW corrections
are presented in Fig. 7. We show the ratio of the leading two-loop
electroweak correction to the
$H\to \gamma \gamma $ decay width (see Eq.(52)) and the full one-loop result
 ($W$- boson plus top contribution).
One notes  that
correction to the decay width is negative and important for $m_H > 500 ~GeV$.
   It is quite clear that the correction is particularly important for
 Higgs
  masses of the order $ 600~GeV$, where there are strong cancellation
 between top
and $W$ contributions to the
leading order result for the $H\gamma\gamma$ vertex.
   This correction blows up at around $m_H \sim 1.5~TeV$. This
general behaviour is quite familiar from previous studies of the large
Higgs mass two-loop radiative corrections \cite{Flei,Dura}.

 Our results can be applied to a more accurate
 estimation of the cross-section
for the reaction $\gamma \gamma \to H \to X$ , which is the basic reaction
for the Higgs boson production at  $\gamma \gamma $-- colliders.
If the Higgs boson is sufficiently heavy, than it is the
broad resonance with a width growing proportionally to $M_H^3$. In this
case our results for the radiative correction to the on-shell value of the
 $H\gamma \gamma$ interaction vertex are not sufficient for the description
of the Higgs shape in this reaction. However, we have given also results
for the
imaginary part of the $H\gamma\gamma$ vertex, hence it is straightforward
to obtain
off-shell value for the $H\gamma\gamma$-vertex,
where the dispersion integrates have to be evaluated
numerically.

\vspace{0.5cm}
\noindent
{\large \bf Acknowledgments:}\\
We like to thank J. Gasser for an informative discussion.

\end{document}